\documentclass[final,3p,times,12pt]{elsarticle} 

\usepackage{amssymb,amsmath}
\usepackage{xspace}
\usepackage{lineno}
\graphicspath{{figs/}}

\journal{Astroparticle Physics}

\def\xmax{\ensuremath{ X_{\rm max}}\xspace}
\def\mxmax{\ensuremath{\langle X_{\rm max} \rangle}\xspace}

\def\nmu{\ensuremath{N_\mu}\xspace}

\def\mlnmu{\ensuremath{\langle \ln N_\mu \rangle}\xspace}

\def\mnmu{\ensuremath{\langle N_\mu \rangle}\xspace}

\begin{document}

\begin{frontmatter}

\title{Sibyll$^{\bigstar}$}

\author[IGFAE]{Felix Riehn}
\author[TW]{Anatoli Fedynitch}
\author[IAP,EKP]{Ralph Engel}

\address[IGFAE]{Instituto Galego de F\'isica de Altas Enerx\'ias (IGFAE),
Universidad de Santiago de Compostela, 15782 Santiago de Compostela, Spain}
\address[TW]{Institute of Physics, Academia Sinica, Taipei City, 11529, Taiwan}
\address[IAP]{Karlsruhe Institute of Technology, Institute for Astroparticle Physics, 76021 Karlsruhe, Germany}
\address[EKP]{Karlsruhe Institute of Technology, Institute of Experimental Particle Physics, 76021 Karlsruhe, Germany}

\begin{abstract}  

\noindent
In the last decade, an increasing number of datasets have revealed a consistent discrepancy between the number of muons measured in ultra-high-energy extensive air showers (EAS) and the numbers predicted by simulations. This gap persists despite incorporating Large Hadron Collider (LHC) data into the tuning of current hadronic interaction models, leading to the phenomenon often termed the ``muon puzzle''. To gain a deeper understanding of the potential origins of this muon puzzle, we have developed Sibyll$^{\bigstar}$, a series of phenomenologically modified versions of Sibyll~2.3d. In these models, we have increased muon production by altering $\rho^0$, baryon--antibaryon pair, or kaon production in hadronic multiparticle production processes. These variants remain within bounds from provided by accelerator measurements, including those from the LHC and fixed-target experiments, notably NA49 and NA61, showing a level of consistency comparable to Sibyll~2.3d. Our findings show that these modifications can increase the muon count in EAS by up to 35\%, while minimally affecting the depth of shower maximum (\xmax) and other shower variables. Additionally, we assess the impact of these modifications on various observables, including inclusive muon and neutrino fluxes and the multiplicities of muon bundles in deep underground and water/ice Cherenkov detectors. We aim for at least one of these model variants to offer a more accurate representation of EAS data at the highest energies, thereby enhancing the quality of Monte Carlo predictions used in training neural networks. This improvement is crucial for achieving more reliable data analyses and interpretations.

\end{abstract}


\end{frontmatter}


\section{Introduction}
\label{s:int}

\noindent
The phenomenon known as the ``muon puzzle'' in extensive air showers (EAS) -- referring to a significant discrepancy between the number of muons observed in EAS with respect to those predicted by standard simulation models -- has led to a critical reassessment of our understanding of hadronic interactions at ultra-high energies~\cite{Albrecht:2021cxw,WHISPicrc2023}. To address this issue, both conventional~\cite{Ostapchenko:2013pia,Drescher:2007hc,Pierog:2006qv,Grieder:1973x1,Baur:2019cpv} and exotic~\cite{Farrar:2013sfa,AlvarezMuniz:2012dd,Rybczynski:2019exi,Anchordoqui:2016oxy,Anchordoqui:2022fpn,Manshanden:2022hgf} modifications of multiparticle production, some including extensions beyond the Standard Model of particle physics, have been proposed. 

While the muon puzzle is of great scientific interest for learning more about hadronic multiparticle production, it also gives rise to large systematic uncertainties in the analysis and interpretation of EAS data. This is especially critical for training machine learning algorithms~\cite{AugerXmaxDNN,PierreAuger:2021nsq,Verpoest:2023qmq,Conceicao:2021xgn} to reconstruct, for example, the depth of shower maximum \xmax and the mass composition of cosmic rays from the ground-level signals of EAS surface arrays. Having a better understanding of the origin of the muon puzzle and being able to simulate air showers with a more realistic muon number are highly desirable for reducing the systematic uncertainties and, in some cases, for the first time, obtaining consistent measurements using different observables.

The typically highly indirect way of deriving information on muon numbers or densities with EAS detectors that are not built for measuring muons separately on the first hand, like the Pierre Auger Observatory~\cite{inclinedReco,Aab:2014pza}, Telescope~Array~\cite{TelescopeArray:2018eph} or IceTop~\cite{IceCubeGeVMuons}, makes a straightforward experimental test of the various proposals to address the muon puzzle very difficult. Detailed simulations of realistic air showers incorporating these proposals, covering different primary particles and a wide range in energy, would have to be compared with existing data sets. However, most of the proposed modifications have either not been worked out to such a level of detail or have not been implemented in a hadronic model capable of generating realistic air showers. 

In this work, we introduce Sibyll$^{\bigstar}$ (in the following also abbreviated as S$^\bigstar$), a set of phenomenological modifications to the Sibyll~2.3d~\cite{sib23eas} hadronic interaction model. These modifications explore various conventional mechanisms with the objective of increasing muon production in EAS. The modified versions of Sibyll can be directly utilized in realistic EAS simulations, ensuring that energy, momentum, and quantum numbers are conserved on an event-by-event basis. One important advantage of these modified versions is that, intentionally, a minimum number of changes has been applied to keep the overall model predictions as much as possible identical to that of Sibyll~2.3d. This will allow a direct comparison with already existing simulations and data analyses/interpretations made with Sibyll~2.3d. Furthermore, making a direct connection between a change of EAS predictions and the underlying modifications to the hadronic interaction model will be possible.

This article is organized as follows. In Sec.~\ref{sec:sibyll-star} the algorithms of making modifications to the particle output of Sibyll~2.3d are introduced and the tuning of the relevant parameters is discussed. It is shown that the predictions of the modified variants of the interaction model are compatible with a selected set of data of accelerator experiments that have direct sensitivity to the modifications. In many cases the agreement of the model predictions with the data even improves. In total, four variants of the Sibyll interaction model are presented. These model variants are then used in Secs.~\ref{sec:EAS} and \ref{sec:inclusive-fluxes} to study the model predictions for air showers and inclusive fluxes, respectively, and to compare them with Sibyll~2.3d as reference. It is argued that, at the level of comparison possible without including a detector simulation, a model variant like, for example, Sibyll$^{\bigstar}$(mix) should properly reproduce the muon data of the Pierre Auger Observatory. The impact of the modifications on the predicted inclusive muon and neutrino fluxes is discussed with emphasis of possible observations with IceCube. Finally, a summary of the results is given in Sec.~\ref{sec:summary}.

\section{Sibyll$^\bigstar$ \label{sec:sibyll-star}}

\subsection{Ad-hoc Event Modification}

\noindent
We construct the custom models by modifying events generated with Sibyll~2.3d. Once the initial event generation is complete, we let all hadronic resonances with lifetimes shorter than that of K$^0_{\rm s}$ decay, except for $\pi^0$. In the next step we replace particles according to the selection probabilities described below.

To replace a particle, for example a $\pi^0$ by a $\rho^0$, in a given generated event without violating energy-momentum conservation one has to change the momenta of (at least) two particles that together reach a sufficiently large invariant mass. Therefore, we iterate through the list of generated particles (event stack) to identify appropriate pairs or triples of pions, while for each pion considering only the five nearest neighboring pions in rapidity. If the sampling criterion for replacing a particle with another one, or replacing both particles, is fulfilled and there is sufficient invariant mass, we replace these pions with a pair of new particles conserving total momentum, invariant mass, and charge. The final momenta are calculated from the invariant mass, the masses of the new particles, and a small transverse momentum component drawn from an exponential distribution in transverse mass. While energy, momentum, and charge are conserved by construction, the (iso)spin conservation is not explicitly maintained\footnote{by favoring $\rho^0$ over the charged $\rho$ states it is in fact explicitly broken}.

The acceptance rate for particle exchanges in our model depends on both the total center-of-mass (c.m.) energy, $\sqrt{s}$, of the interaction and the fractional longitudinal momentum of the considered secondary particle $x_{\rm F} \equiv p_{z}/p_{z,\mathrm{max}}$ (with momenta $p$ in the c.m.\ frame). The exchange probability is parameterized by
\begin{equation}
    P_i ~=~ P_{i,0} \cdot |x_{\rm F}|^{\epsilon_i} \cdot f(\sqrt{s}, E_{\rm thr}) \ .
    \label{eq:enh-par}
  \end{equation}
Whether the modifications are applied mainly to forward or central particle production depends on the chosen value for the exponent $\epsilon_i$ in the $x_{\rm F}$-dependence. If $\epsilon_i=0$, all particles receive equal weight, effectively preserving the shape of the original distribution in longitudinal phase space. In the case of $\epsilon_i = 1$, the modification is proportional to $x_{\rm F}$ and the forward part of the $x_{\rm F}$-spectrum undergoes significant enhancement (with maximal enhancement at $x_{\rm F}\to 1$). And so on for larger values of $\epsilon_i$.

The energy dependence of $f(\sqrt{s},E_{\rm thr})$ follows a logarithmic form. It is specifically chosen to ensure that the modification probability is precisely zero below a set threshold energy, $E_{\rm thr}$, and reaches its nominal value of $P_{0,i}$ at lab energies of $10^{19}\,$eV ($1.37\times10^5\,$GeV in the c.m.\ frame). This parameterization of the energy dependence ensures a gradual change in particle production. At low energies, where fixed-target experiments effectively constrain particle production across the entire phase space, no event modification is permitted. At LHC energies, only the central region is well-constrained by measurements, leaving room for modifications in particle production for the forward phase space. The constraints from the LHCf experiment~\cite{LHCf:2008lfy,Tiberio:2023xgj,Piparo:2023dgd,LHCf:2020hjf} only apply to forward neutral particle production, whereas FASER~\cite{FASER:2023zcr} and SND~\cite{SNDLHC:2023pun} or future experiments at the FPF~\cite{Feng:2022inv} may provide stricter constraints on forward charged particle production by observing muons and neutrinos~\cite{Kling:2021gos}. Finally, at the UHECR energy scale, where laboratory experiments offer no constraints, particle production undergoes substantial modification. 

As an alternative, we implement a more significant increase in the exchange rate at energies above $13\,$TeV in the c.m.\ frame, expressed as $ P_{i} \to P_{i,0} + P_{i,\mathrm{HE}} \cdot f_{\rm HE}(\sqrt{s}=13\,\mathrm{TeV})$. This mode simulates a rapid transition to new physics beyond the c.m.\ energy of the LHC.

We employ this algorithm for all interactions in an EAS if one can assume the enhancement should be universal by its nature. Only in the case of $\rho^0$ enhancement, directly related to the leading particle effect of pions,
we restrict it solely to pion-air interactions.

\subsection{Tuning Enhancement Parameters with Accelerator Measurements}

\begin{figure}[tb]
  \centering
  \includegraphics[width=\textwidth]{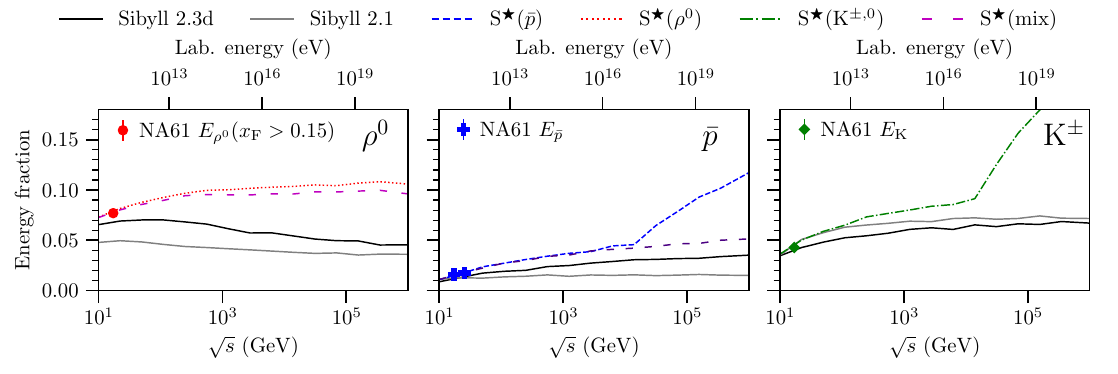}
  \hfill
  \includegraphics[width=\textwidth]{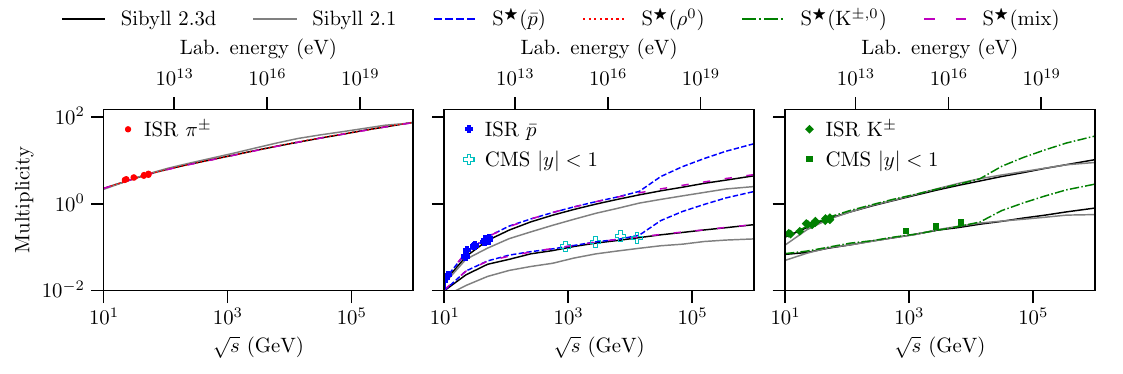}
  \caption{Upper panel: fraction of projectile energy carried by $\rho^0$ (left), anti protons (center) and charged kaons (right) in $\pi^-$C collisions~\cite{Aduszkiewicz:2017anm,NA61SHINE:2022tiz}. The values of the energy fraction in kaons were obtained by integrating the published $x_{\rm F}$-spectra. Lower panels: multiplicities of $\pi^+$, anti protons and charged kaons in pp collisions~\cite{Sirunyan:2017zmn,Chatrchyan:2012qb,Albini:1975iu}. Lines are Sibyll~2.3d, Sibyll~2.1 and different variants of Sibyll$^\bigstar$.}
  \label{fig:en-and-mult-na61}
\end{figure}

\noindent
Leveraging the algorithm described above, we construct distinct variants of Sibyll~2.3d with the explicit goal of increasing muon production in EAS. Within an EAS, the bulk of muons arise from meson decay, which in turn are predominantly generated through the interactions of other hadrons (the hadronic cascade). The total number of muons in a shower, therefore, depends critically on the energy retained within the hadronic cascade at each interaction step. More precisely, increasing the fraction of energy ($f_{\rm had}$) carried by hadrons that reinteract in the shower (which, to a very good approximation, encompasses all hadrons except neutral pions and $\eta$ resonances) leads to a corresponding increase in the total muon count within the shower. The important point here is that the hadronic energy is changed across the whole hadronic cascade, from the highest energy interactions down to the low energy interactions that produce the pions that will then decay into muons. This way small changes in each interaction accumulate to produce a large effect in the muons at the ground. The opposite approach, dramatically changing the highest energy interaction and leaving the low energy interactions unchanged, also serves to increase the number of muons at the ground but would affect the fluctuations in the number of muons as well. These, however, were observed to be described well by current models~\cite{PierreAuger:2021qsd}. 

We select three specific modifications recognized for their efficacy in increasing the hadronic energy and enhancing muon production: $\rho^0$ production, baryon--antibaryon pair production, and kaon production enhancement~\cite{Ostapchenko:2013pia,Drescher:2007hc,Pierog:2006qv,Grieder:1973x1,Anchordoqui:2022fpn}. We denote these variants as S$^\bigstar$($\rho^0$), S$^\bigstar$($\bar{p}$), and S$^\bigstar$(K$^{\pm,0}$).

In the $\rho^0$ variant, $\pi^0$ are directly substituted with $\rho^0$. For the baryon pair and kaon pair variants, charge-neutral combinations of two or three pions are replaced with  $p\bar{p}$ or $n\bar{n}$ pairs, and K$^+$K$^-$ or K$^0$ $\bar{\mathrm{K}}^0$ pairs respectively. We adjust the parameters for each variant to ensure a sufficiently good description of laboratory measurements. Representative examples include the energy fractions in $\rho^0$, antiprotons, and kaons measured by the NA61 experiment~\cite{NA61SHINE:2022tiz,Aduszkiewicz:2017anm} (upper panels in Fig.~\ref{fig:en-and-mult-na61}), as well as the collection of multiplicity measurements of pions, antiprotons, and kaons in proton-proton interactions (lower panels in Fig.~\ref{fig:en-and-mult-na61}~\cite{Sirunyan:2017zmn,Chatrchyan:2012qb,Albini:1975iu}). Sibyll~2.3d does not always describe the lab.\ data well. For example in comparison with the energy fractions in $\rho^0$, $\bar{p}$ and $K^\pm$ measured in the NA61 experiment (c.m.\ energy around $17\,$GeV), the predictions by Sibyll~2.3d are consistently lower than the observation (see upper panels in Fig.~\ref{fig:en-and-mult-na61}). To allow the improved description of these data we set the threshold energy $E_{\rm thr}$ in Eq.~\eqref{eq:enh-par} to $5\,$GeV in the c.m.\ frame.

\begin{figure}[tb]
  \centering
  \includegraphics[width=\textwidth]{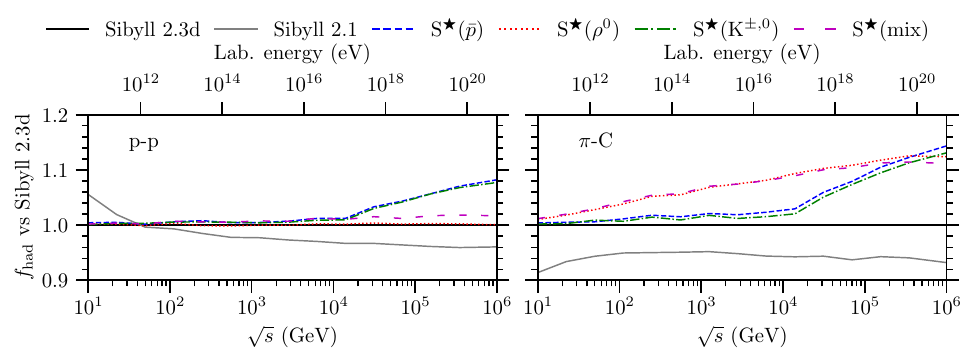}
  \caption{Change in the hadronic energy fraction between Sibyll~2.3d and the enhanced variants. In the left panel pp interactions are shown. In the panel on the right $\pi$C interactions are shown. }
  \label{fig:had-energy-frachtion-change}
\end{figure}

Within the phase space extending beyond the scope of laboratory measurements, we set the parameters to achieve an approximately equal hadronic energy fraction ($f_{\rm had}$) among the different variants at primary energies around $10^{19}\,$eV. We target a value of $\approx 0.82$, an increase of $\approx 10\%$ over Sibyll~2.3d (see Fig.~\ref{fig:had-energy-frachtion-change}). $\rho^0$ production has a strong impact on muon production in EAS by redirecting the energy from the electromagnetic channel into hadrons through the substitution of $\pi^0$ by $\rho^0$ in particular for forward $\pi^0$. However, the data only support an enhancement of forward $\rho$ production for meson projectiles~\cite{Adamus:1986ta,Agababyan:1990df,AguilarBenitez:1991yy} and not for protons. Consequently, we apply the modifications in the $\rho^0$ variant exclusively for the interaction of pion projectiles. In contrast, the modifications introduced for baryon pair and kaon production encompass all projectile types. The availability of proton-proton measurements at significantly higher energies (LHC energy scale) compared to measurements with pion projectiles constrains the baryon pair and kaon variants to much higher energies than the $\rho^0$ variant. This requires a more rapid increase of the modifications in these two variants at energies beyond the LHC to attain the targeted hadronic energy fraction at ultra high energies.

We construct a fourth variant, S$^\bigstar$(mix), that combines both $\rho^0$ and baryon pair production enhancements. In this scenario, a more moderate increase in $\rho$ production and no rapid increase in baryon production at high energies are needed to achieve similar effects on the showers. Table~\ref{tab:variants} presents a detailed overview of the parameters employed in the different variants.

\begin{table}[h]
    \caption{Parameters in different variants of Sibyll$^\bigstar$. $\rho$-mix and $\bar{p}$-mix refer to the subcomponents of the \textit{mixed} variant. \label{tab:variants}}
    \begin{center}
      \renewcommand{\arraystretch}{1.1}
      \begin{tabular}{ccccc}
        \hline
        Label & $P_{i,0}$ & forward weight $\epsilon_i$ & projectiles & $P_{i,\mathrm{HE}}$ \\     
        \hline
        $\rho^0$            & 0.9 & 0.3 & mesons & -  \\
        $\bar{p}$           & 0.5 & 0.7 & all & 0.25  \\
        $K^{\pm,0}$           & 0.5 & 0.8 & all & 0.3  \\
        $\rho$-mix          & 0.95 & 0.4 & mesons & -  \\
        $\bar{p}$-mix       & 0.5 & 0.7 & all & -  \\    
        \hline
      \end{tabular}
    \end{center}
\end{table}

\begin{figure}[tb]
  \centering
  \includegraphics[width=\textwidth]{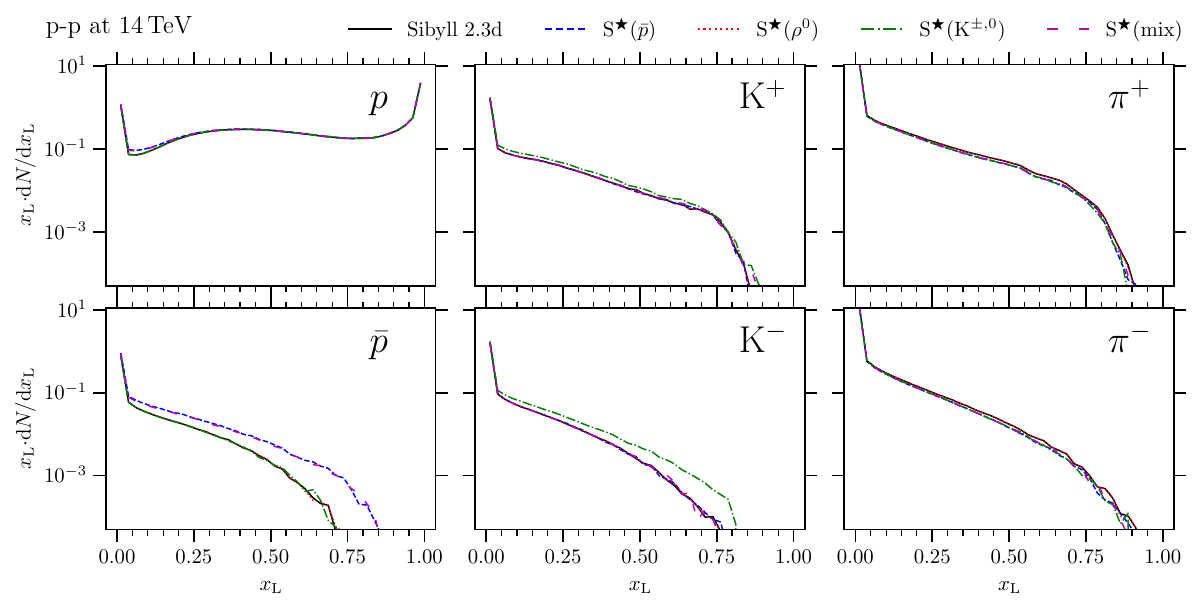}
  \caption{Spectra of the energy fraction in the lab.\ frame for pions, kaons and protons in pp interactions at $14\,$TeV for Sibyll~2.3d and different variants of S$^\bigstar$.}
  \label{fig:xl-spectra-14tev-pp}
\end{figure}

The above algorithm introduces only gradual changes both to the overall production rate (and energy) of the particles but also in the shape of the respective production spectra. The effect on the spectra at $14\,$TeV is demonstrated in Fig.\ref{fig:xl-spectra-14tev-pp} for the case of pp interactions. In the figure the spectra of the energy fraction in the lab.\ frame ($x_{\rm L}$) for pions, kaons and protons are shown. The spectra are weighted by $x_{\rm L}$. Both baryon and kaon enhancement lead to substantial changes in the production spectra at large $x_{\rm L}$ in the forward region which leads to the overall increase in the fraction of energy carried by hadrons (see Fig.\ref{fig:had-energy-frachtion-change}). As the forward phasespace is outside the coverage of the LHC experiments the observed production rates (Fig.~\ref{fig:en-and-mult-na61}) are not affected.

\section{Predictions for Extensive Air Showers \label{sec:EAS}}

\noindent
In the following, we explore the impact of the presented Sibyll variants on muon production predictions in EAS. First, we review the muon production in air showers and later discuss the impact on specific observables. All predictions were calculated using CORSIKA~v7.7420~\cite{Heck98a} with FLUKA~2021.2.9~\cite{fluka2014} as the low-energy interaction model. 

\subsection{Expectation from the Heitler-Matthews Model}
As indicated before, muons in EAS are produced by the decay of mesons in the hadronic cascade. The larger the hadronic cascade (the more mesons) the more muons will be produced in a shower. The size of the hadronic cascade is determined by the energy of the primary particle and the fraction of energy that is emitted into photons in each interaction\footnote{A small fraction of mesons and subsequently muons also arises from the electromagnetic cascade via photon-nucleus interactions.}. The more energy is retained in the hadronic cascade, the greater the meson abundance and subsequent muon production. For a proton primary in a shower with energy $E_0$, the average number of muons, $\langle N_\mu^p(E_0) \rangle$, is given by
\begin{equation}
  \langle N^p_\mu(E_0) \rangle = \left( \frac{E_0}{\varepsilon} \right)^\beta.
  \label{eq:avg-nmu-p}
\end{equation}
Here, $\beta$ depends on the energy transfer from the hadronic to the electromagnetic shower component, and $\varepsilon$ represents the critical energy – the point where mesons decay rather than interact, effectively terminating the cascade. In the Heitler-Matthews model~\cite{Matthews:2005sd}, assuming identical particle yields and only pion production in each hadronic interaction, $\beta$ relates to the ratio of the logarithms of the multiplicity of neutral pions (decaying into two photons) to all pions. This simplified case assumes the cascade ends around $130\,$GeV, where pion decay becomes more likely than interaction. However, real EAS involve various hadron species with differing critical energies, influencing both $\beta$ and $\varepsilon$. In simulations with only pions, $\beta$ is around 0.88, while full Monte Carlo simulations yield values closer to 0.9~\cite{Engel:2011zzb}.

For nuclear primaries, the Heitler-Matthews model utilizes the superposition principle. This assumes an EAS initiated by a nucleus with $A$ nucleons is equivalent to $A$ individual proton-initiated showers with energy $E_0/A$. Consequently, the average number of muons for nuclear primaries, $\langle N_\mu^A(E_0) \rangle$, can be written as
\begin{equation}
  \langle N^A_\mu(E_0) \rangle = A \cdot \langle N^p_\mu(E_0/A) \rangle = A^{1-\beta} \cdot \langle N^p_\mu(E_0) \rangle
  \label{eq:superpos}
\end{equation}
\noindent This dependence implies a shrinking separation between different mass primaries as $\beta$ and the overall muon count increase.

The various Sibyll variants aim to enhance muon production through distinct mechanisms, primarily by increasing energy retention within the hadronic cascade (effectively raising $\beta$):
\begin{itemize}
\item $\rho^0$ production: Directly substitutes $\pi^0$ with $\rho^0$ (decaying into charged pions), altering the charged/neutral pion ratio but not the critical energy or overall balance of hadron species.

\item Baryon pair enhancement: Due to baryon number conservation, a baryon cascade terminates only when the baryons become non-relativistic (around $1\,$GeV). Creating more baryons within the hadronic cascade lowers the critical energy and significantly enhances the number of low-energy muons produced in the final stage of the shower.
  
\item Kaon enhancement: Increases production of charged and neutral kaons, leading to a higher critical energy since kaons have a shorted lifetime than pions ($\varepsilon \sim 1\,$TeV). This increases $\beta$ and muon count, but with slightly higher-energy muons.
\end{itemize}
Analyzing the specific predictions of these enhanced models in terms of muon production and their impacts on EAS observables will be the focus of the next sections.

\subsection{Muons at Production}\label{sec:muon-production}

\begin{figure}[tb]
    \centering
    \includegraphics[width=0.48\textwidth]{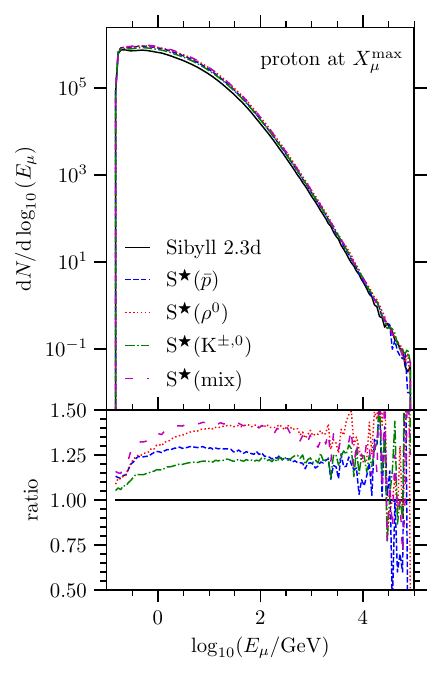}
    \hfill
    \includegraphics[width=0.48\textwidth]{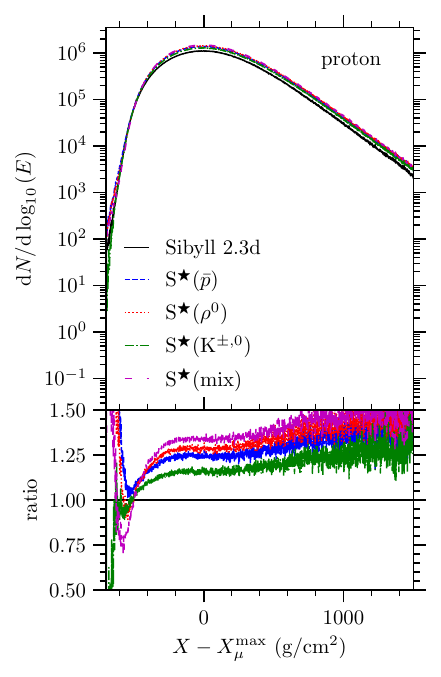}
    \caption{\label{fig:mu-prod}
    Left panel: Energy spectrum of muons produced at the shower maximum. Right panel: Distribution of muon production as a function of depth (relative to the depth of maximum production). Both figures show the average over 100 proton showers with a zenith angle of $67^\circ$ and a primary energy of $10\,$EeV.}
\end{figure}

The zenith angle of the primary, the distance of the observer from the shower axis (core distance), the observation height, and even atmospheric conditions (temperature, season) can slightly influence the number of muons detectable at a specific point on the surface. However, given the energy, transverse momentum, and depth distribution of muons emerging from the hadronic cascade, local muon yields can be reliably calculated~\cite{Cazon:2012ti,Cazon:2022msf}. Therefore, analyzing the distributions of muons at their production point is crucial for understanding the origin of the muon puzzle.

The left panel of Fig.~\ref{fig:mu-prod} presents the energy spectrum of muons produced around the shower maximum. The bottom of the figure shows the ratios with respect to Sibyll~2.3d, revealing the differences between the variants. While $\rho^0$ and kaon enhancements increase high-energy muons, baryon enhancement is more effective at lower energies. All S$^\bigstar$ variants exhibit a similar change in the profile of muon production (shown on the right in Fig.~\ref{fig:mu-prod}). The position of the production maximum ($X_\mu^{\rm max}$) remains largely unaffected (less than $5\,$g$/$cm$^2$ change, not shown here).

\subsection{Energy Dependence}\label{sec:energy-dependence}

\begin{figure}[tb]
\centering
\includegraphics[width=1.0\textwidth]{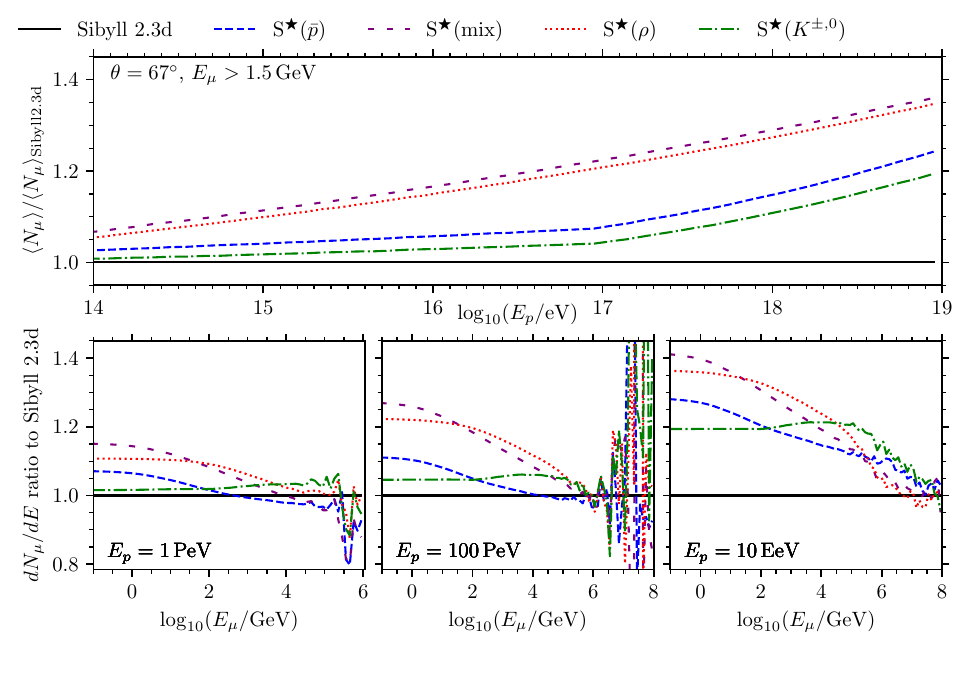}
\caption{\label{fig:nmu-energy-scaling-and-spectrum}
Top Panel: Energy dependence of the surface muon number as a function of cosmic ray energy. The expectation for nuclei can be obtained through the superposition model in Eq.~(\ref{eq:superpos}). The trend mirrors that of the hadronic energy fraction (Fig.~\ref{fig:had-energy-frachtion-change}). Bottom Panels: Muon energy spectra at three representative primary cosmic ray energies. Despite the distinct primary energies, the qualitative features of the muon spectra remain remarkably consistent. In the high-energy limit $E_\mu/E_p \rightarrow 1$, dominated by the first proton-air interaction, the different S$^\bigstar$ models produce nearly identical results.}
\end{figure}

Figure~\ref{fig:nmu-energy-scaling-and-spectrum} illustrates the energy dependence of the muon multiplicity, following the trend indicated by the hadronic energy fraction (compare with Fig.~\ref{fig:had-energy-frachtion-change}). The flat ratios of the energy spectra in the bottom panels indicate that the relative increase of the muon number compared to the Sibyll~2.3d reference is independent of energy thresholds (up to a few tens of GeV). However, at higher energy thresholds, such as those encountered in underground muon studies or muon bundle analyses, the S$^\bigstar$ variants may begin to demonstrate more significant differences.

\subsection{Inclined Showers at the Pierre Auger Observatory}\label{sec:auger-inclined}

\begin{figure}[tb]
  \centering    
  \includegraphics[width=0.7\textwidth]{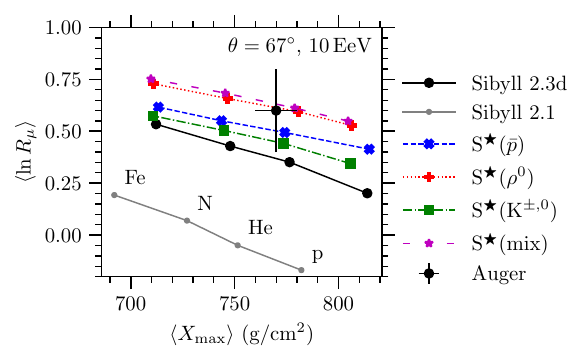}
  \caption{Comparison of the predicted values for \mxmax and \mnmu from various Sibyll$^\bigstar$ variants to the measurements obtained from the Pierre~Auger~Observatory~\cite{PierreAuger:2021qsd}.}
  \label{fig:rmu-vs-xmax}
\end{figure}

Measurements of the average number of muons in inclined showers at the Pierre~Auger~Observatory have unveiled a deficit in the muon count predicted by air shower simulations~\cite{Aab:2014pza,PierreAuger:2021qsd}. This deficit becomes particularly apparent when the depth of shower maximum (\mxmax) is also considered. Since both \mxmax and \mlnmu are proportional to $\langle \ln{A} \rangle$, the model predictions would fall along a line in the \mxmax-\mlnmu plane for any given composition scenario. The slope of this line, as dictated by Eq.~\eqref{eq:superpos}, is proportional to the exponent $\beta$. The muon deficit in simulations is revealed by the fact that the Auger data point is well off the predicted lines for any model. Assuming a composition inferred from \mxmax measurements, the deficit for Sibyll~2.3d is around 26\%~\cite{Bellido:2017cgf}.

In Figure~\ref{fig:rmu-vs-xmax} the predicted values for \mxmax and \mnmu from the S$^\bigstar$ variants and the measurements obtained from the Pierre~Auger~Observatory are shown. Both the $\rho^0$ (red line) and mixed variants (magenta line) demonstrate a sufficient increase in muon number to describe the Auger data. The baryon pair (blue line) and kaon enhanced variants (green line) also lead to a substantial increase in muon count, but not to the extent required to fully describe the data. It is worth noting that the slope $\beta$ increases for all the S$^\bigstar$ variants. While it is around 0.93 for Sibyll~2.3d, it rises to a value closer to 0.95 for the S$^\bigstar$ variants.

\begin{figure}[tb]
  \centering
  \includegraphics[width=0.48\textwidth]{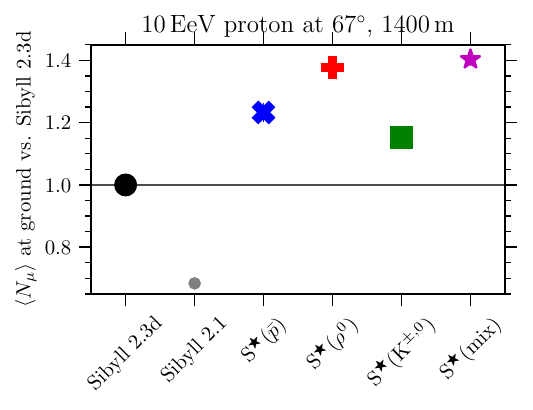}
  \hfill
  \includegraphics[width=0.48\textwidth]{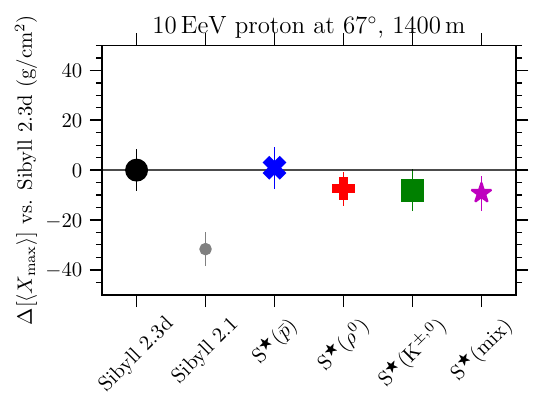}
  \vfill
  \includegraphics[width=0.48\textwidth]{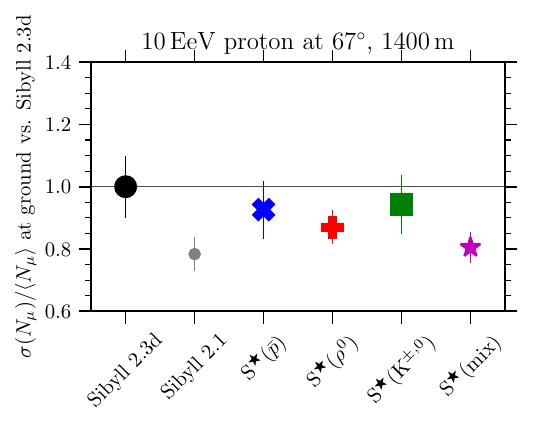}
  \hfill
  \includegraphics[width=0.48\textwidth]{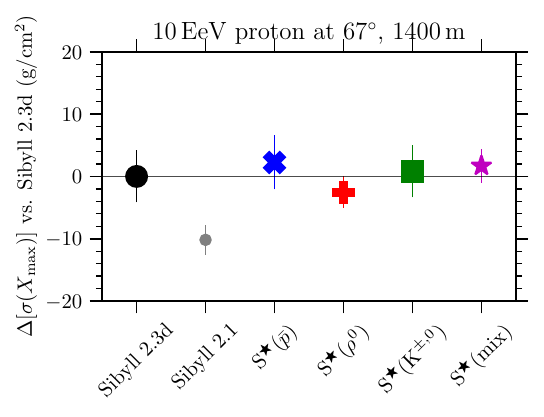}

  \caption{\nmu and \xmax for proton showers at $67^\circ$ across Sibyll variants. For \nmu (left panel) there is a substantial increase in the average, up to 35\%, for the mixed and $\rho$ variant. The gray line represents the required increase in muon count to align with the data for a mixed composition seen by the Pierre~Auger~Observatory. The variation in the average of the shower maximum (panel at the top on the right) between Sibyll~2.3d and its variants is less than $7\,$g$/$cm$^2$. The fluctuations (lower panels) decrease by up to 20\% for \nmu (left) but are not affected for \xmax (right). }
  \label{fig:nmu-xmax-change}
\end{figure}

In Figure~\ref{fig:nmu-xmax-change} the changes in the average (top) and the fluctuations (bottom) in \nmu (left) and \xmax (right) across S$^\bigstar$ variants are shown for proton showers at a zenith angle of 67$^\circ$. The predicted value of \mxmax, shown in the right panel on the top, is almost unaltered due to the changes in muon production. At $10\,$EeV the differences between all S$^\bigstar$ variants and Sibyll~2.3d remain below $7\,$g$/$cm$^2$. In the top left panel of Fig.~\ref{fig:nmu-xmax-change} the relative change in the predicted number of muons is shown. It is important to note that due to the change in $\beta$ to $\beta + \delta$ (with $\delta>0$) and the dependence of the muon count on $A^{1-\beta}$ in the superposition model (see Eq.~\eqref{eq:superpos}), a $c$-fold increase in muon count for proton showers translates to only a $c \, A^{-\delta}$-fold increase for a shower of mass number $A$. Since the primary composition in the Auger data is not pure protons ($A=1$), the necessary increase in muon count to describe the data (denoted by the gray line in Fig.~\ref{fig:nmu-xmax-change}) is approximately 40\% for pure proton showers.

Finally, it is worth emphasizing that the shower-to-shower fluctuations in \xmax (shown in the lower panel on the right) remain unaffected by the enhancements implemented in the S$^\bigstar$ variants. Fluctuations in the number of muons (shown in the panel on the lower left) on the other hand decrease by approximately 20\% in case of S$^\bigstar$(mix) and around 10\% for S$^\bigstar$($\rho^0$). For the pure baryon and kaon enhanced variants the fluctuations decrease only slightly.

\subsection{Vertical Showers at Sea Level: Lateral Distribution}

Most cosmic ray experiments primarily detect vertical cosmic rays. Ground-based detectors achieve mass composition sensitivity by separating the muonic and electromagnetic signal components~\cite{Aab:2016vlz}. As mentioned previously, the relative contributions of these components reaching the ground depend on the location of the experiment. As a generic case study, we examine the impact of the S$^\bigstar$ variants on the muon component for vertical showers reaching a sea-level experiment in the US standard atmosphere.

\begin{figure}[tb]
  \centering
  \includegraphics[width=\textwidth]{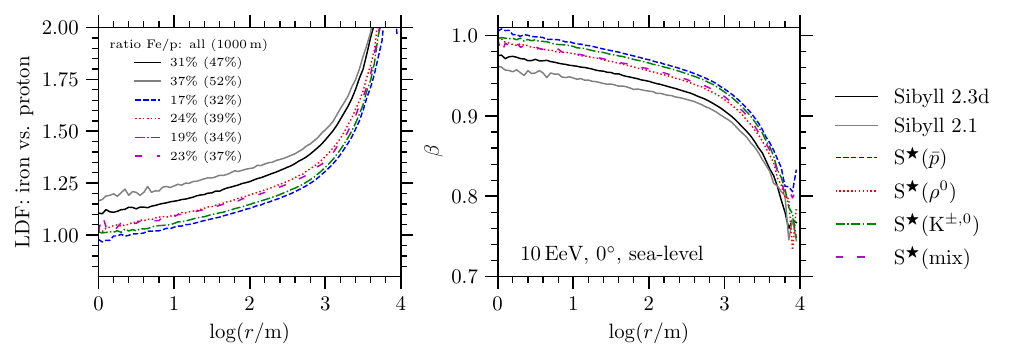}
  \caption{
    Ratio between the average number of muons in iron and proton showers (left panel) and the slope $\beta$ (right panel), both as functions of the distance from the shower axis. The inset numbers in the left panel show the total muon ratio between iron and proton showers (first number) and the muon density ratio at $1000\,$m (second number in parentheses).
  }
  \label{fig:beta-vs-r}
\end{figure}

According to the superposition model (Eq.~\eqref{eq:superpos}), the ratio between the total number of muons in a shower of a primary with $A$ nucleons and a proton shower is $A^{1-\beta}$. Since enhancing muon production via hadronic particle production requires an increase in the hadronic energy (leading to larger $\beta$), this inevitably reduces the mass resolution that is the difference between proton and iron induced showers. For Sibyll~2.3d with $\beta$ around 0.93, the separation between proton and iron in the number of muons is close to 30\%. For the S$^\bigstar$ variants, it can be as low as 17\% (see legend in Fig.~\ref{fig:beta-vs-r}). Note that this connection between muon production and mass resolution does not apply to modifications affecting only the first high-energy interactions since such changes would not influence the development of the hadronic cascade ($\beta$), only its initial conditions. However, this scenario is constrained or even excluded by: 1) Current model predictions matching the fluctuations in the number of muons observed at the Pierre~Auger~Observatory~\cite{PierreAuger:2021qsd}; 2) The muon deficit being observed across several orders of magnitude in energy~\cite{WHISPicrc2023}. That is to say estimates of the mass resolution based on the total number of muons with models that produce less muons than are seen in the data are likely an overestimation.

The situation may not be as dramatic for specific experiments, however. As the different variants alter the muon energy spectrum, the lateral distribution of muons reaching the ground will change slightly. Notably, $\beta$ or the mass separation will vary with the distance from the core. This effect is shown in Fig.~\ref{fig:beta-vs-r}. At around $1000\,$m, the separation for Sibyll~2.3d is already around 50\%, and for the variants, it increases to around 30\% (with a correspondingly lower $\beta$). The reason for this difference as a function of lateral distance is that different phases of the shower development dominate at different distances~\cite{sib23eas,Cazon:2012ti,Cazon:2022msf,MULLER2018174,Maris:2009uc}.

\begin{figure}[tb]
  \centering
  \includegraphics[width=0.48\textwidth]{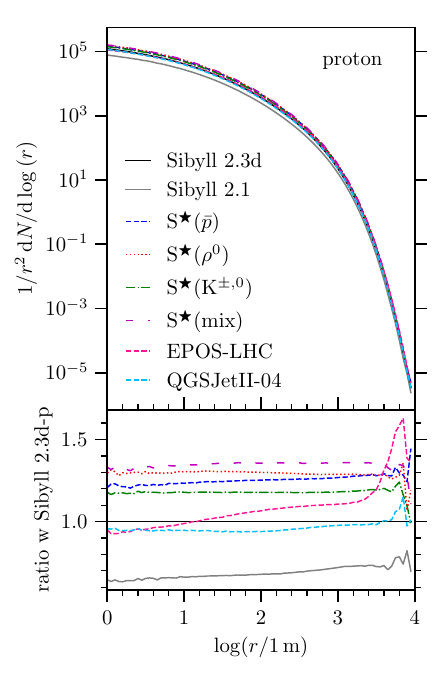}
  \hfill
    \includegraphics[width=0.48\textwidth]{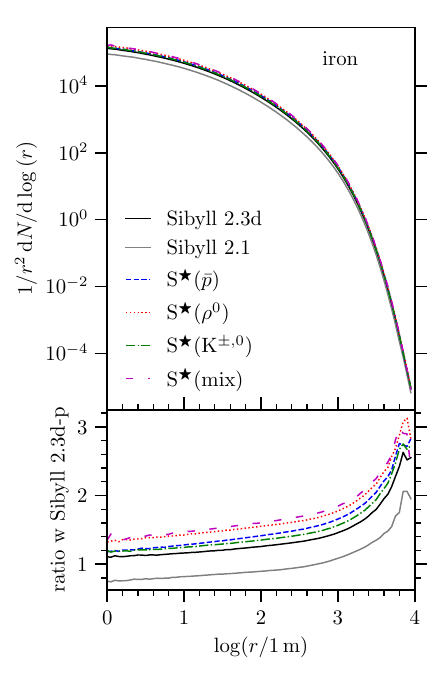}
  \caption{Lateral distribution of muons at the ground. Proton primaries are shown in the left panel, iron primaries are shown on the right. The ratios in the lower panels are with respect to proton primaries for Sibyll~2.3d. Between the variants of S$^{\bigstar}$ and Sibyll~2.3d there is only a slight variation in the shape of the LDF. The largest difference in the shape is between Sibyll~2.3d and EPOS-LHC.
  }
  \label{fig:LDF}
\end{figure}

In Fig.~\ref{fig:LDF} the lateral distribution of muons and the ratio of the distributions with regard to Sibyll~2.3d are shown for proton and iron primaries. The largest difference in shape (slope) are between Sibyll~2.3d and EPOS-LHC~\cite{Pierog:2013ria}. The variants of S$^{\bigstar}$ and QGSJet-II~04~\cite{Ostapchenko:2010vb} are all remarkably similar in shape to Sibyll~2.3d. The largest difference between the models is by far in the absolute number of muons in the relevant lateral distance range of about $300$ to $3000$\,m.

\subsection{Vertical Showers at the South Pole}

The lowest-energy measurement of the muon deficit comes from IceTop, the surface detector of the IceCube neutrino observatory, measuring muon densities from $3\,$PeV to $100\,$PeV~\cite{IceCubeGeVMuons,Verpoest:2023qmq}.

\begin{figure}[tb]
  \centering
  \includegraphics[width=0.48\textwidth]{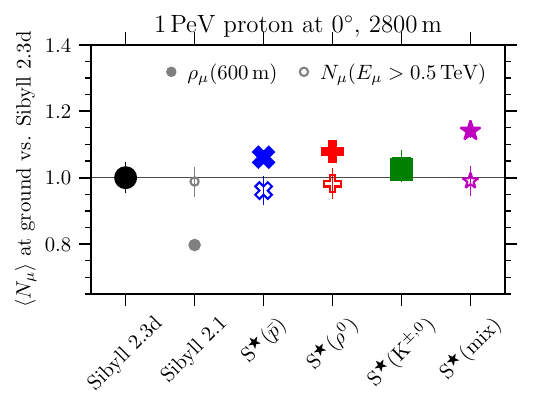}
  \hfill
  \includegraphics[width=0.48\textwidth]{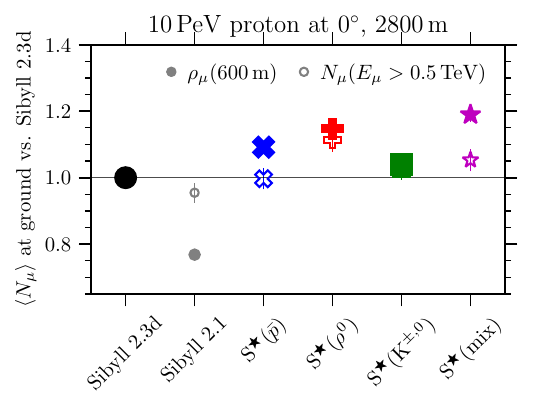}
  \caption{
    Ratio of muon counts for vertical proton showers at the South Pole for different S$^\bigstar$ variants relative to Sibyll~2.3d. Full symbols are muon densities at $600\,$m lateral distance (surface muons). Empty symbols are total muon counts with energies above $0.5\,$TeV (in-ice muons). The left panel corresponds to $1\,$PeV primary protons, and the right panel to $10\,$PeV primary protons.
    \label{fig:nmu-icetop}
    }
\end{figure}

Figure~\ref{fig:nmu-icetop} presents the ratio of muon counts for the S$^\bigstar$ variants relative to Sibyll~2.3d for vertical proton showers at the South Pole. Full symbols represent muon densities at $600\,$m lateral distance (surface muons), while empty symbols indicate total muon counts with energies above $0.5\,$TeV (in-ice muons). 

For S$^\bigstar$ variants that align with Auger data at $10\,$EeV (the $\rho^0$ and mixed variants), surface muons at $1\,$PeV increase by 10-15\%, while in-ice muons remain unchanged. The situation differs at $10\,$PeV. Here, the $\rho^0$ variant exhibits a roughly 15\% increase in both surface and in-ice muons, while the mixed variant shows a 20\% increase in surface muons but only a 5\% increase in in-ice muons.

The analysis by the IceCube collaboration found that Sibyll~2.1 is consistent with both surface and in-ice muon data, assuming the mass composition from the Global Spline Fit (GSF) cosmic ray flux model~\cite{IceCubeGeVMuons,Verpoest:2023qmq,Dembinski:2017zsh}. However, Sibyll~2.3d predicts a roughly 20\% increase in surface muons for proton primaries compared to Sibyll~2.1. While this increase is smaller for a mixed composition, it is unlikely that Sibyll~2.3d describes the surface data. For in-ice muons, Sibyll~2.1 and Sibyll~2.3d predict similar numbers. The S$^\bigstar$ variants, with a 30-40\% increase in surface muons relative to Sibyll~2.1, are even less likely to describe the surface and in-ice data consistently. Note however that the effect of snow and ice on the muons at the South Pole is complex so our definition of in-ice and surface muons only approximately corresponds to the quantities measured by IceCube.

\section{Predictions for Inclusive Fluxes and Underground Muons \label{sec:inclusive-fluxes}}

\noindent
In this section, we study what implications the S$^\bigstar$ variants have for inclusive atmospheric muon and neutrino flux calculations, as well as muon bundles observed in deep underground and Cherenkov detectors in water or ice. Calculations were performed using the MCEq code~\cite{Fedynitch:2015zma,sib23flux} with DPMJET-III-19.3~\cite{AFthesis} as the low-energy interaction model, potentially leading to slight variations compared to CORSIKA calculations. The Global Spline Fit (GSF) primary flux model~\cite{Dembinski:2017zsh} is used throughout this work to emphasize differences induced by S$^\bigstar$ variants.

Sibyll~2.3d is extensively used for the modeling of high-energy atmospheric neutrino fluxes in IceCube~\cite{IceCube:2016rnb}. Except for an inconsistency in the seasonal variations of the neutrino flux~\cite{IceCube:2023qem}, no significant deviations have been observed. Since inclusive flux predictions depend equally on the cosmic ray flux model, interpreting deviations from measured spectra cannot be unequivocally attributed to the hadronic model alone. Stronger model discrimination might be achieved by combining multiple observables, including muon and neutrino spectra, angular distributions, deep underground muon rates, muon bundle multiplicities, and seasonal variations, potentially shedding light on the origin of the muon excess.

\subsection{Inclusive Fluxes}
\begin{figure}[tb]
    \centering
    \includegraphics[width=0.9\textwidth]{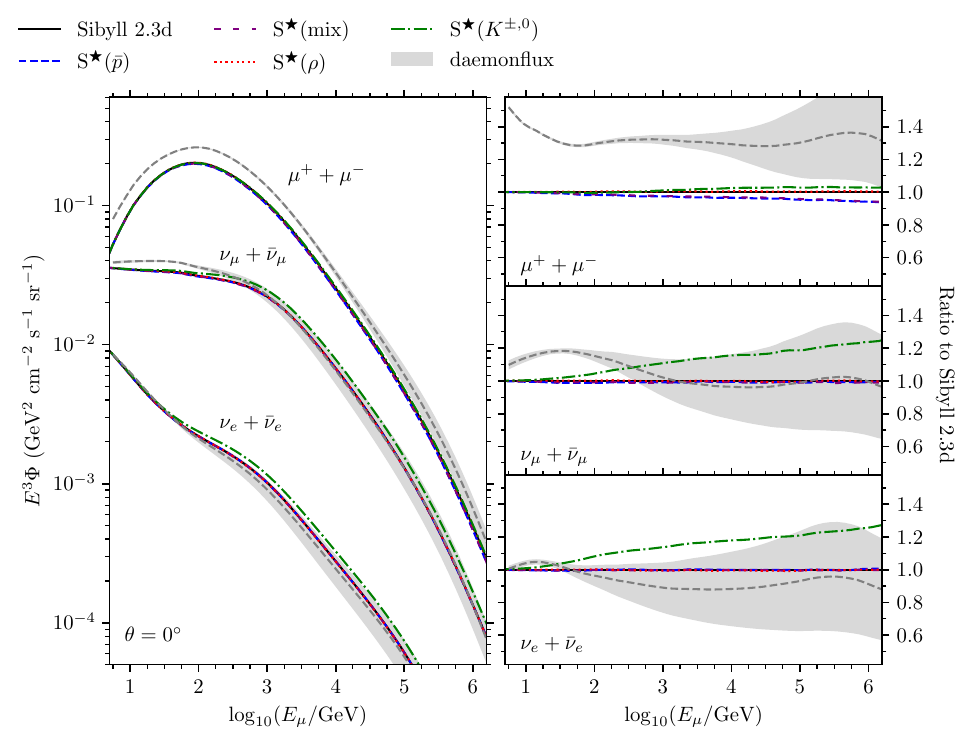}
    \caption{Inclusive muon and neutrino fluxes calculated for a vertical zenith angle. These are conventional fluxes predominantly originating from decays of charged pions and kaons. Notably, S$^\bigstar$ modifications have a negligible impact, except for the S$^\bigstar$(K$^{\pm,0}$) model's enhancement of neutrino fluxes due to increased kaon decays. Calculations at inclined zenith angles yield qualitatively similar results with minor differences. The data-driven daemonflux model~\cite{Yanez:2023lsy}, derived from accelerator, cosmic-ray, and surface muon data, is included for comparison, with error estimates extrapolated to higher energies.
        \label{fig:incl_flux_vertical}
        }
\end{figure}
\begin{figure}[tb]
    \centering
    \includegraphics[width=0.8\textwidth]{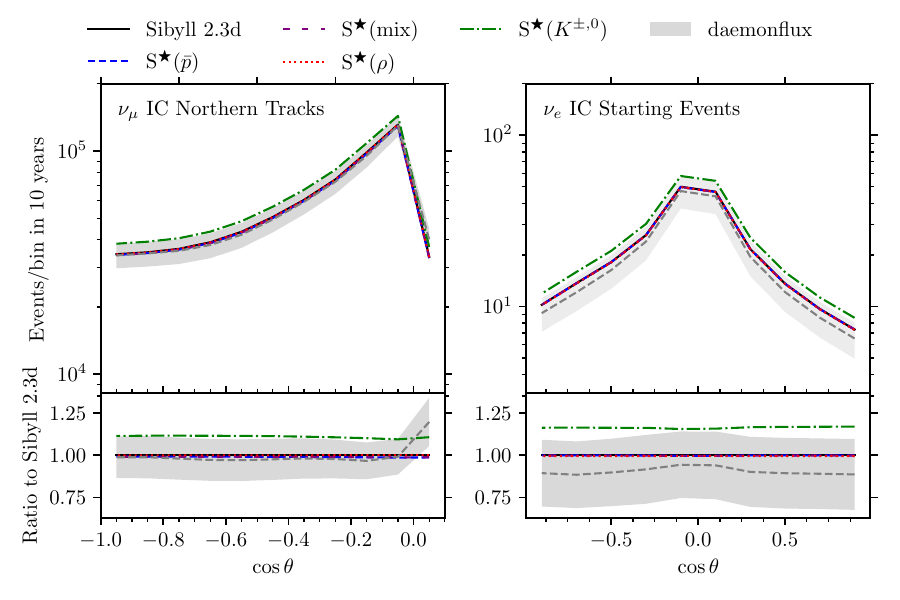}
    \caption{Calculated zenith distributions of tracks (mostly muon neutrino charged-current events) and electron neutrino events contained within the instrumented volume of IceCube, using effective areas from public data releases~\cite{IceCube:2015qii,IceCube:2014rwe} scaled to 10 years of livetime. No specific zenith dependence is observed between S$^\bigstar$ versions, while daemonflux predicts more events from the horizon due to increased neutrinos from muon decay. As expected from Fig.~\ref{fig:incl_flux_vertical}, the S$^\bigstar$(K$^{\pm,0}$) model predicts 15-20\% higher neutrino rates.
    \label{fig:zenith_distribution_icecube}
        }
\end{figure}

The prediction for the inclusive spectra of muons, muon neutrinos and electron neutrinos (and their respective anti-particles) are shown in Fig.~\ref{fig:incl_flux_vertical}. Apart from the S$^\bigstar$(K$^{\pm,0}$) model, the predictions from the S$^\bigstar$ variants demonstrate no significant difference compared to Sibyll~2.3d. To some extent this result is expected, since inclusive fluxes are dominated by the decay of (charged) pions and kaons produced in the first interaction and less so by particles produced in the sub-cascades as it is the case in air showers. Therefore S$^\bigstar$($\rho^0$) and S$^\bigstar$($\bar{p}$), which only lead to small changes in the production of pions in individual interactions at these energies, show no effect. In S$^\bigstar$(K$^{\pm,0}$), the enhanced strangeness production in proton-air interactions leads to more kaon parents decaying into neutrinos, but with little effect on muon fluxes since these, at the energies shown here, predominantly originate from pions. The gray line and band in Fig.~\ref{fig:incl_flux_vertical} is the prediction by a data-driven model, the so-called daemonflux. This model is fit to various muon measurements performed at the surface, taking into account systematic uncertainties and constraints from lab.\ measurements~\cite{Yanez:2023lsy}. In the energy range from $10\,$GeV to $1\,$TeV it represents the ``world data'' on inclusive fluxes to well within 10\%. The muon fluxes in all Sibyll models (and most other interaction models~\cite{Fedynitch:2018cbl}) are about 30\% lower than the daemonflux model that is than the experimentally observed flux of muons. A preliminary KM3NeT Collaboration study observes a similar 30\% deviation between the underwater reconstructed muon rate and Sibyll~2.3d~\cite{Romanov:2023jti}. A too low muon prediction is also supported by comparisons with underground laboratory data~\cite{Fedynitch:2021ima}. However, daemonflux only incorporates surface muon data up to a few TeV and predictions in the hundreds TeV range are affected by large extrapolation errors. Whether the discrepancy observed here in the inclusive flux of muons at low energies is related to the muon excess observed in air showers at ultra-high energies is not evident. While both require an increased production of pions to achieve consistency between data and predictions, they are sensitive to different of parts of the atmospheric cascades. For one, as shown in Fig.~\ref{fig:incl_flux_vertical}, the S$^\bigstar$ models, which do solve the muon puzzle at ultra-high energies, are not the answer to the atmospheric muon puzzle.

For both neutrino flavors, all Sibyll variants and the daemonflux model agree well within errors.
A similar observation holds for the zenith distribution of predicted neutrino events in IceCube (Fig.~\ref{fig:zenith_distribution_icecube}). No significant deviation from the zenith distribution predicted by Sibyll~2.3d is observed, except for a 15-20\% higher neutrino rate in the S$^\bigstar$(K$^{\pm,0}$) case. Some of the IceCube $\nu_\mu$-focused analyses~\cite{IceCube:2016umi,IceCube:2020tka} have hinted at a higher atmospheric neutrino flux normalization, however no specific measurement has been performed. The main difference from the daemonflux model is observed at the horizon, where the muon production distance is far enough from the surface for decay into neutrinos even at TeV energies to occur, so the higher muon flux estimate propagates into neutrinos.

\subsection{Underground Muons}

\begin{figure}[tb]
  \centering
  \includegraphics[width=.57\textwidth]{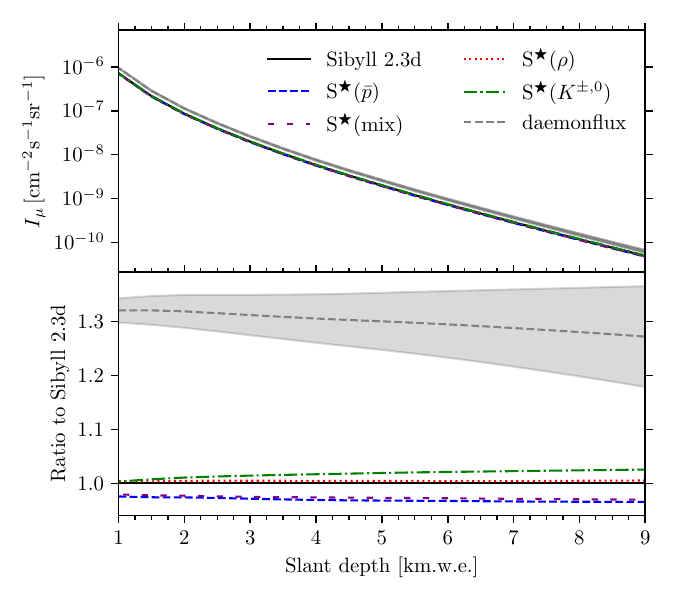}
  \caption{Vertical equivalent muon fluxes in water, obtained using the MUTE code, which combines MCEq and PROPOSAL codes for underground muon flux calculations~\cite{Fedynitch:2021ima}.
    \label{fig:vertical_equivalent_fluxes}
  }
\end{figure}
  
\begin{figure}[tb]
  \centering
  \includegraphics[width=1\textwidth]{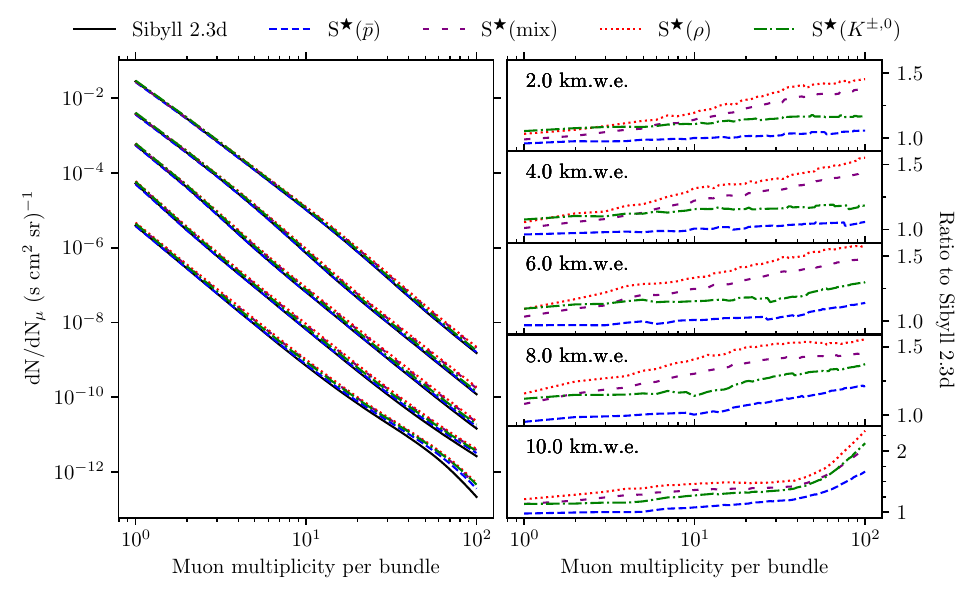}
  \caption{Muon bundle multiplicity distribution in water for different slant depths. The zenith angle at the surface has been accordingly adjusted as $\theta = \arccos(1{\rm km.w.e.}/X)$. The distribution of muon bundle multiplicities (left panel) strongly depends on the primary cosmic ray flux model. When scaled to a reference model (here Sibyll~2.3d, see right panels), differences between the S$^\bigstar$ emerge, which become more pronounced in the rate of very high multiplicity bundles with $\sim100$ muons.
    \label{fig:muon_bundle_muliplicities}
  }
\end{figure}

Muons observed in laboratories deep underground, in water, or in ice exhibit significant energy loss. The overburden acts as a high-pass filter, selectively absorbing muons with lower energies ($<1\,$TeV). This makes them valuable probes for the high-energy portion of the surface spectrum, potentially distinguishing between different mechanisms that enhance muon numbers in S$^\bigstar$ variants. Figure~\ref{fig:vertical_equivalent_fluxes} presents calculated vertical equivalent muon fluxes in water, obtained using the MUTE code, which combines MCEq and PROPOSAL~\cite{Dunsch:2018nsc} codes for underground muon flux calculations~\cite{Fedynitch:2021ima}. The vertical equivalent intensity refers to the rate of single muons above a low energy threshold. Due to median muon energies being in the hundreds of GeV range, the threshold has minimal impact. While S$^\bigstar$ models exhibit more variation here than at the surface, the changes are still too subtle to be discerned given uncertainties in interaction and cosmic ray flux models. As expected from surface flux comparisons (Fig.~\ref{fig:incl_flux_vertical}), the data-driven daemonflux model predicts a 20-30\% higher flux, resembling the recent preliminary results from KM3NeT's ORCA and ARCA detectors~\cite{Romanov:2023jti}.

Figure~\ref{fig:muon_bundle_muliplicities} shows distributions of average muon bundle multiplicities, representing the muonic cores of air showers that reach specific depths underwater or in rock. Muon bundle multiplicity distributions strongly depend on the primary cosmic ray spectrum and mass composition, particularly around the energy range of the CR knee~\cite{MACRO:2002jmi}. However, with better composition constraints from other measurements, ratios between low and high-multiplicity events at fixed zenith angles (or slant depths) could serve as a proxy to discriminate between S$^\bigstar$ models, complementing surface muon multiplicities. At large depths and/or inclined zenith angles the differences between S$^\bigstar$ and the default Sibyll~2.3d reach up to 50\% or more at near-horizontal incidence angles.  
  
While muon bundles can be studied in ice~\cite{IceCube:2015wro}, water-based detectors like KM3NeT~\cite{KM3Net:2016zxf} and Baikal-GVD~\cite{Avrorin:2022lyk} might offer superior multiplicity separation due to a larger fraction of direct Cherenkov light. Upcoming, technologically advanced, large-volume neutrino telescopes, such as P-ONE~\cite{P-ONE:2020ljt} and TRIDENT~\cite{2023NatAs.tmp..211Y}, are anticipated to significantly enhance bundle multiplicity measurements. Furthermore, large scale radio arrays in the future may provide the opportunity to directly constrain the flux of atmospheric muons at PeV energies~\cite{Pyras:2023crm}. 

\section{Summary \label{sec:summary}}

The new Sibyll$^\bigstar$ models are a set of phenomenological modifications to the well-known Sibyll~2.3d hadronic interaction model. They aim to increase the number of muons in extensive air showers (EAS) to address the "muon puzzle." These modifications increase the fraction of energy retained in the hadronic cascade, leading to more meson production and, consequently, more muons. At the same time, predictions for other air shower observables are intentionally not changed, or changes kept to a minimum.

The models are exploring different conventional mechanisms for increasing muon production: enhanced $\rho^0$ production, increased baryon--antibaryon pair production, and increased kaon production. A mixed variant is integrating both $\rho^0$ and baryon pair production enhancements as a baseline approach for comparison to data. The different variants have been adjusted to describe the data on particle production that are available from accelerator experiments, while also achieving a desired increase in the hadronic energy fraction at ultra-high energies.

We observe that the S$^\bigstar$ variants can increase the number of muons in EAS by up to 35\%, depending on the variant and shower configuration. This occurs with a very small impact on the average depth of shower maximum \mxmax, typically less than $7\,$g$/$cm$^2$. The $\rho^0$ and mixed variants can provide a sufficient increase in muon number to match measurements from the Pierre Auger Observatory, but are unlikely to improve compatibility with IceCube muon data.

Notably, modifications that address the muon deficit imply a change in the slope parameter $\beta$ that describes the energy dependence of the muon number. This is understood within the Heitler-Matthews model and confirmed by our numerical studies. A direct consequence is a reduced sensitivity of the muon number of showers as mass sensitive observable. However, a good mass separation capability can be recovered at larger lateral distances from the shower core.

Regarding inclusive muon and neutrino fluxes, the S$^\bigstar$ modifications have little effect on inclusive muon and neutrino energy spectra compared to Sibyll 2.3d. The only exception is the S$^\bigstar$(K$^{\pm,0})$ variant, which increases neutrino fluxes due to the enhanced kaon production and subsequent decays. The predicted zenith distributions of tracks and cascades in IceCube show no specific dependence on S$^\bigstar$ versions. Therefore, given that the IceCube observations are consistent with the prediction by Sibyll~2.3d, the predictions by all the S$^\bigstar$ variants will similarly be consistent.

While the changes in the surface flux are subtle, the S$^\bigstar$ variants exhibit larger variations in the underground muon fluxes, especially at high depths and inclined zenith angles. S$^\bigstar$ models predict different distributions of muon bundle multiplicities in water or ice, offering a potential discriminator between them. The next generation large-volume neutrino telescopes could significantly improve measurements of these multiplicities.

All-in-all, the S$^\bigstar$ models provide a promising framework for addressing the muon puzzle in EAS. It allows a more consistent description of EAS observables, for high energy cosmic rays and neutrino observatories. By representing accurately the correlations between observables in the same event and for different events, in a coherent way conserving energy-momentum and most quantum-numbers, it allows the derivation of full MC sets to train machine learning algorithms. The different variants of S$^\bigstar$ induce similar changes for the number of muons and the shower maximum but can be distinguished by other measurements.

The code is available as part of the {\sc chromo} package\footnote{https://github.com/impy-project/chromo}~\cite{Dembinski:2023esa}. A standalone version is available upon request from one of the authors (FR).

{\small
  \paragraph{Acknowledgements}
  The authors acknowledge many fruitful discussions with colleagues of the Pierre~Auger and IceCube Collaborations. We are grateful to Jenni Adams, Sofia Andringa and Ruben Concei\c{c}\~{a}o for comments on the manuscript of this article.
  FR and RE are supported in part by the European Union’s Horizon~2020 research and innovation programme under the Marie~Sk\l{}odowska-Curie grant agreement No.{} 101065027 and by BMBF grant No.{} 05A2023VK4, respectively. The authors jointly acknowledge the valuable computing resources provided by the Academia Sinica Grid Computing Center (ASGC), supported by the Institute of Physics at Academia Sinica. AF is supported by the Academia Sinica Grand Challenge Seed Program Grant AS-GCS-113-M04.

{\small
  \paragraph{Dedication}
The authors dedicate this work to their late colleague Thomas K.\ Gaisser, with whom they have enjoyed many years of fruitful collaboration on developing the hadronic interaction model Sibyll and on calculating predictions for astroparticle physics applications.
  
\paragraph{Declaration of generative AI and AI-assisted technologies in the writing
process}
Generative AI has been used to improve text flow and grammar of parts of the existing text.
}

\bibliographystyle{elsarticle-num}

\end{document}